# Roger of Hereford: the twelfth-century astronomer who put Hereford on the map, literally

Richard de Grijs
(Macquarie University, Sydney, Australia)

## Islamic southern Spain

In the Arab[1] world, the period spanning the seventh to thirteenth centuries is commonly known as the Golden Age of Islam. At a time when much of Europe was languishing in an environment dominated by religious intolerance and wars—with progress in science, technology and innovation largely stagnant—cultural, economic and scientific developments flourished in the history of Islam.

It took until the last few centuries of the Middle Ages before Arabic innovations found their way into Europe, mainly through North Africa and Islamic southern Spain. By the middle of the twelfth century, the slow disintegration of Muslim rule in southern Spain had enabled development of a tentative dialogue between the Latin and Islamic worlds.[2]

Numerous Arabic manuscripts on natural philosophy, many of them astrological in nature, were translated into Latin for the first time. Many had originally been translated into Arabic, in the eighth and ninth centuries, from original source material in Greek, Persian or Hindi,[3] including Ptolemy's masterpieces *Geographia*[4] and *Almagest*. In addition, an increasing appetite for travel—for both pilgrimages and the Crusades—fostered a renewed interest in geography.[5]

By the twelfth century, Toledo, in southern Spain, had developed into a centre of Islamic science, becoming an important nexus in Islamic–Latin exchanges. The city was home to Gerard of Cremona (ca. 1114–1187), for instance, a well-known Italian translator of Arabic manuscripts, including Ptolemy's *Almagest* (translated 1175). Daniel of Morley (ca. 1140–ca. 1210), English natural philosopher and astronomer, made it abundantly clear that he was keen to move from Paris—which he felt was dominated by 'law and pretentious ignorance'—to Toledo[6] to rub shoulders with the world's leading philosophers. Northern European theological teachings were still based on traditional authorities, whereas the intellectual environment in southern Spain centred on 'reason' and calculations to attain an improved understanding of a range of natural phenomena.

Daniel of Morley composed his *Philosophia*—also known as his *Liber de naturis inferiorum et superiorum*—specifically to educate Bishop John of Norwich (fl. 1175–1200) in the teachings of Toledo (see Figure 1, left). His astronomical chapters were heavily influenced by Arabic sources, including the ninth-century writings of Abū al-ʿAbbās Aḥmad ibn Muḥammad ibn Kathīr al-Farghānī (see Figure 1, right). Other English natural philosophers and astronomers, such as Adelard of Bath (ca. 1080?–ca. 1142–1152?) and Robert of Ketton (fl. 1141–1157), had similarly brought back to England new insights gleaned from the Arabic philosophy and astronomical practices they had encountered in Spain. Adelard, in particular, had undertaken his own translations of Muḥammad ibn Mūsā al-Khwārizmī's (ca. 780–ca. 850) astronomical tables (see below). He is also known to have translated an abbreviated version of Abu

Ma'shar al-Balkhi's (787–886) astrological treatise *Introductorium Maius* (1120). This represented new material to the scholars and natural philosophers of northwestern Europe, allowing them to calculate and predict planetary positions for the first time and, thus, engage in practical astrology, as had meanwhile become the norm in the Arab world.[7]

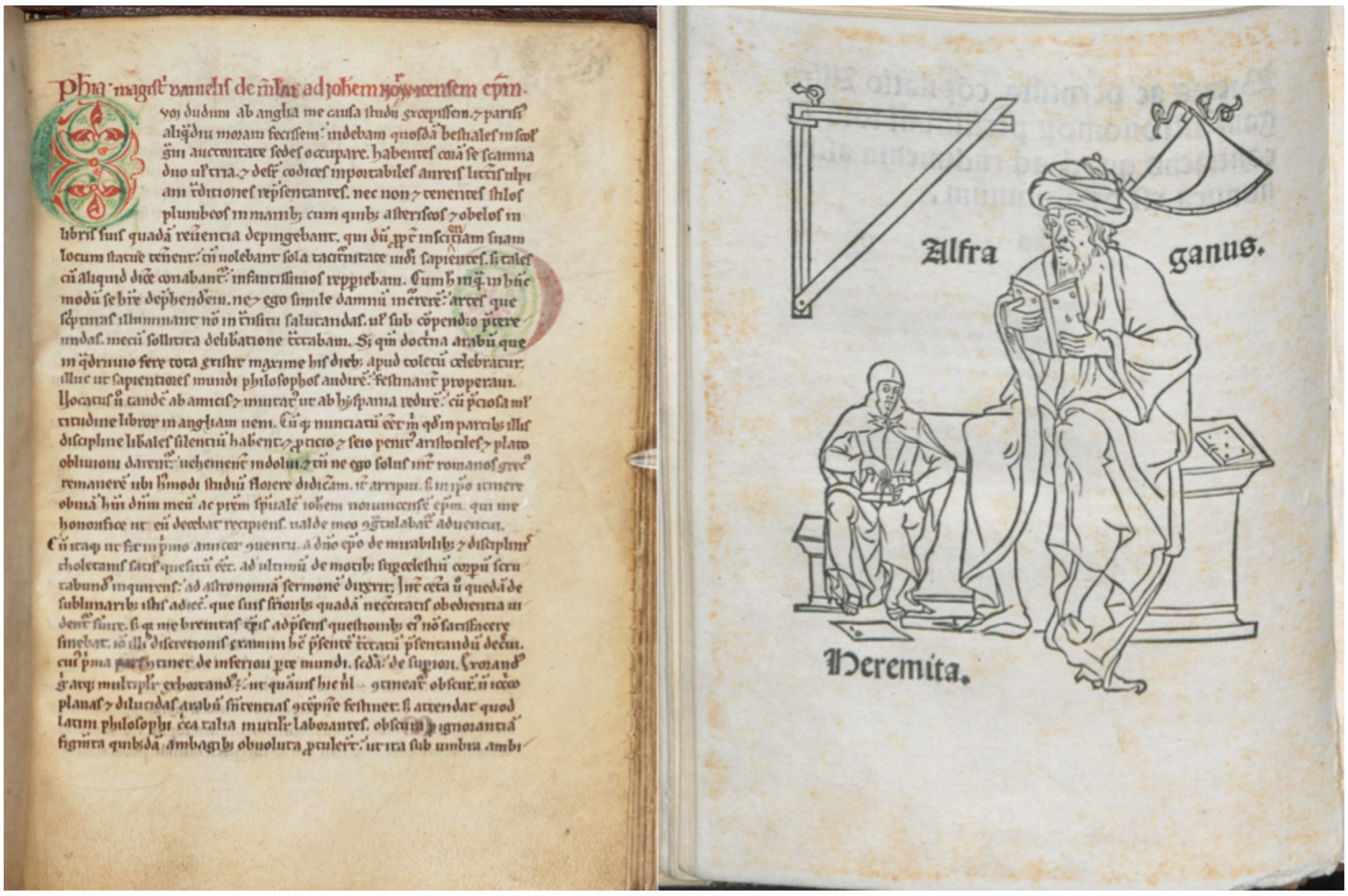

**Figure 1.** (*left*) Daniel of Morley's *Philosophia*, Arundel MS 377, f.88r (Courtesy: British Library, via Wikimedia Commons; public domain); (*right*) Abū al-ʿAbbās Aḥmad ibn Muḥammad ibn Kathīr al-Farghānī, *Compilatio Astronomica* (Andrea Belfort, 1493). (Courtesy: BEIC Digital Library, via Wikimedia Commons; public domain)

## Practical astrology

For practising Muslims it was important to pray at least five times a day while facing in the direction of Mecca. Mosques, hence, had to be oriented such that praying in the correct direction was easily achieved. In turn, that required one to know the geography of Mecca in a regional context. In addition, Islamic scholars were interested in practical astronomy and astrology from a belief that the positions of the Sun, the Moon, the planets and the stars above a given location would determine one's destiny. And so, being able to predict future conjunctions of celestial bodies would presumably offer insights into the course of human destiny. However, for this approach to be more generally useful, it was imperative that the astronomical tables calculated for a given location on Earth could be adapted to any other location.[8]

Europeans, on the other hand, were more interested in applying their newly acquired expertise in astronomical position determination to predict future (astrological) events.[9] Translations of Arabic astrological and mathematical manuscripts, compilations of astronomical tables for northern European parallels and meridians (latitudes and longitudes), and promotion of the use of astrolabes for practical astronomy thus all contributed to the overarching goal of enabling predictions of the future. Observations using astrolabes and the mathematical manipulation of astronomical tables allowed for accurate predictions of the positions of celestial bodies. This complemented astrological texts which, in turn, supposedly provided insights into the deeper meaning of the observations.[10]

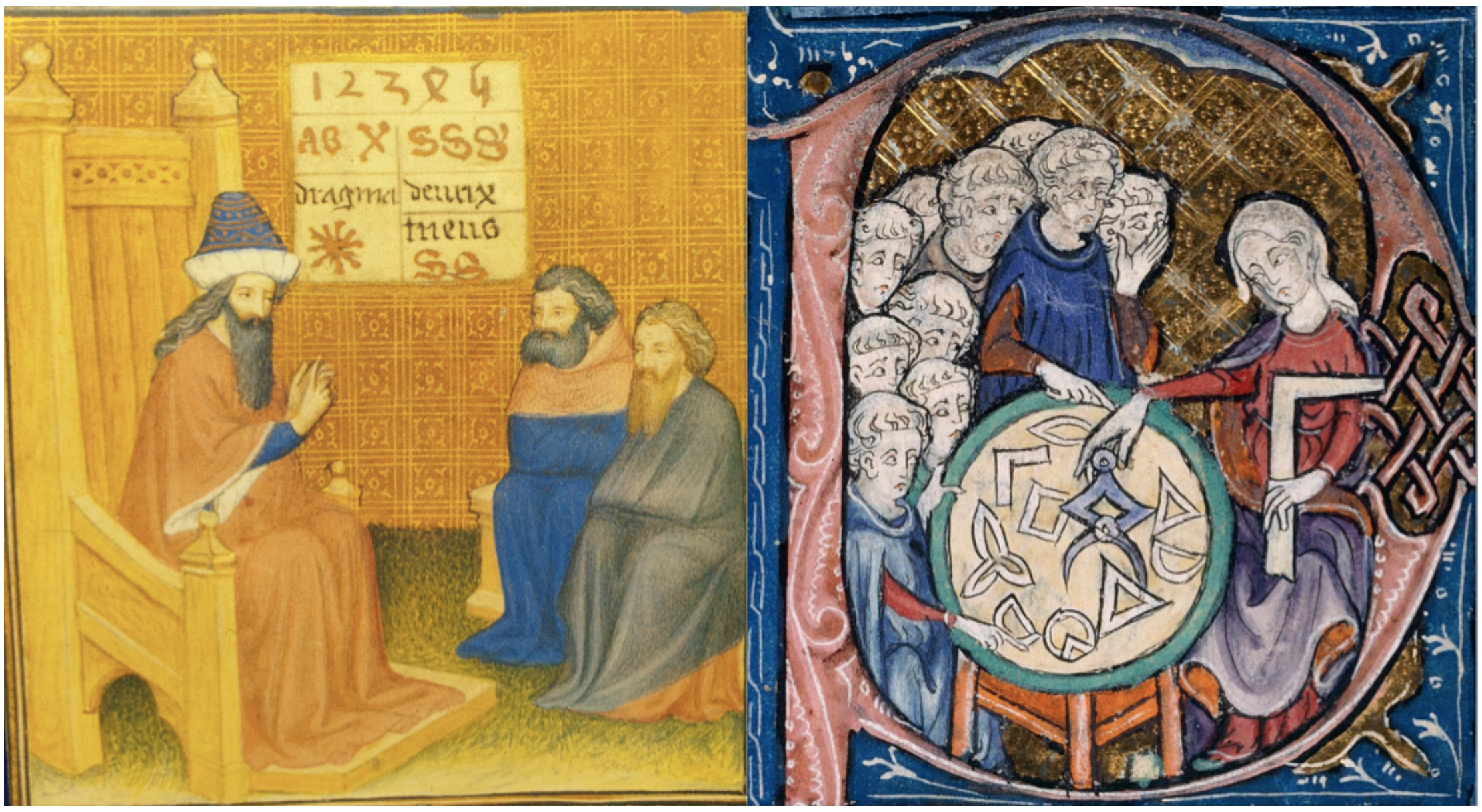

**Figure 2.** (*left*) Adelard of Bath, teaching (*Adelardi Bathensis Regulae abaci SCA 1*, ca. 1400) (Courtesy: Leiden University, via Wikimedia Commons; Creative Commons Attribution-Share Alike 4.0 International license); (*right*) Illustration on the frontispiece of Euclid's *Elements*, in the translation attributed to Adelard of Bath, Burney MS 275, f.293r. (Courtesy: British Library, via Wikimedia Commons; public domain)

England naturally developed into a significant centre of the new learning in western Europe.[11] In addition to pioneers like Adelard of Bath (Figure 2) and Daniel of Morley, a new generation of scholars (specifically, abacists and computists), whose work was characterised by explicit adoption of Arabic influences, flourished across England and the continent. Eleventh-century pioneers included Turchillus Compotista (Thurkil),[12] Philippe de Thaon (Thaun),[13] Bede 'the Venerable',[14] Nimrod 'the Astronomer',[15] Helperic of Auxerre[16] and Gerland.[17] The latter scholar was an eleventh-century mathematician best known for his *Computus*,[18] a type of manuscript dedicated to calculating the exact dates on which important events in the church calendar, such as Easter, are due to occur.[19] Whereas astronomy dealt with all observational phenomena involving the planets in our solar system, *computus* was concerned only with the study of lunar and solar motions across the sky:

> *Compotus est scientia distinctionis temporum secundum motum duorum principalium planetarum, solis videlicet et lune*. (*Computus* is the science of temporal reckoning according to the motions of the two principal 'planets', namely the Sun and the Moon.)[20]

Numerous English scholars sought inspiration in the Arabic teachings promulgated in southern Spain. Meanwhile, educated Jewish scholars frequented the main centres of learning in England,[21] particularly the secular cathedral schools.

## Hereford's cathedral school

Hereford, and specifically its cathedral school,[22] played a particularly important role in the transition of English scholarship to the new learning. The cathedral city was an important border town in the twelfth century, controlling some of the main routes between England and Wales. Manuscripts on Arabic science, the astrolabe, the abacus and 'rhythmomancy'—an approach to teach mathematics—had most likely found their way to the English West Country and southwest England through connections with the School of Lotharingia (Lorraine, France)[23] or through Jewish sources.[24] Contacts between the West Country and Lotharingia predate the eleventh-century Norman Conquest, however.[25]

Bishop Robert of Lorraine (also known as Robert Losings or Robert 'the Lotharingian'; d. 1095) had been educated at the cathedral school in Liège, which was famous for its teaching of mathematics. He is thought to have introduced the abacus to England. An astronomer of some repute and renowned for a computistic treatise advocating a renumbering of the years,[26] he arrived in Hereford in 1079. His arrival generated keen local interest in scholarship associated with the *computus*.

The cathedral school at Hereford flourished under Bishop Robert's guidance, with a succession of Masters insisting that even theology had much to learn from astrology and astronomy.[27] As a case in point, in the preface to his *Tractatus de Computo*[28] (1176), Roger of Hereford (fl. 1172?–1198?) asserted that theology, 'which is about the knowledge of the creator', has made use of astronomy as a means to provide context to everyday life.[29]

Hereford acquired a reputation for philosophy, scholarship and scientific learning. The city became known as a centre for the applied sciences, with expertise that went well beyond the use of the abacus, the astrolabe and computational methods applied to the compilation of astronomical tables. Its scholars' focus extended to the deeper, new mathematics of axiom and structure inspired by the translation of Euclid's *Elements*.

The virtues of this elevated scholarly and philosophical environment are extolled by a poem from about 1195–1197.[30] It was addressed to Gerald of Wales (ca. 1146–ca. 1223), historian and archdeacon of Brecon, by the poet and Anglo–Norman canon (cleric) Simon de Freine (de Fresne). It was a clear attempt to entice Gerald to move to Hereford (although the cleric did not take the bait):

| | |
|---|---|
| Urbs Herefordensis multum tibi competit, in qua | The city of Hereford stands with you, in which |
| Proprius est trivii quadriviique locus, | Is the true place of the trivium and the quadrivium, |
| Floruit et floret, in hac specialiter urbe | It has flourished and is flourishing, and in this city particularly |
| Artis septenae praedominatur honos.[31] | The honour of the seven arts rules foremost.[32] |

Here, the 'seven arts' refer to the seven liberal arts, that is, the *trivium* arts of rhetoric, grammar and dialectic (logic), combined with the *quadrivium* subjects astronomy, mathematics, geometry and music. In addition, astrology (which appears to largely have resembled conventional astronomy), geomancy (divination based on patterns), 'fysis' or 'fisis' (which may refer to natural philosophy or the study of medicine[33]) and law are also explicitly mentioned in the poem. These subjects were complemented with the teaching of theology, as we learn from the introduction to Roger of Hereford's *Computus*.

The curriculum taught at the Hereford cathedral school closely resembled that of the universities, the *studia generalia*, that would eventually emerge in late-medieval England—including the seven liberal arts, philosophy and at least one of the higher faculties of theology, law and medicine. The cathedral school at Hereford appears to have been a pioneering institution that had carved out a niche between the traditional religious schools and the much larger secular schools that would eventually develop into universities.[34]

**Roger of Hereford**

Thus, in the second half of the twelfth century an atmosphere of general excitement about the potential of the 'new science' prevailed in the West Country. In response, Hereford cathedral developed into a focal point for high-level scholarship, attracting leading scholars such as Walter Map (1130–ca. 1210), Roger of Hereford, Gerald of Wales, Robert Grosseteste (ca. 1168–1253) and Bishop William de Vere (d. 1198).

Roger of Hereford stands out among his peers as an enlightened scholar who made more practical use than most of the full astronomical and astrological knowledge base available in England in the late twelfth century. A significant body of recent scholarship focuses on twelfth-century ecclesiastical developments, including those relating to Robert Grosseteste[35] and to Roger of Hereford's *Computus* (concisely summarised here). However, much less scholarly emphasis is placed on Roger's astronomical calculations, particularly those which allowed him to establish an important reference meridian at Hereford. I specifically aim at doing the latter in this paper.

Perhaps surprisingly, Roger's precise identity remains somewhat obscure. Known in Latin as Rogerus Hereford(i)ensis, Henofortensis or Herefordus,[36] his alternative identity as Roger Infans originates from the title, *Prefatio magistri Rogeri Infantis in compotum*, of his *Computus* and calendar of 1176.[37] The name Roger Infans also appears in a number of charters at Hereford dating from between 1172 and 1198.

In addition, a gloss (a marginal annotation) in a work by the twelfth-century translator and philosopher Alfredus Anglicus (also known as Alfred 'the Englishman' or Alfred of Sareshel) refers to him as Rogerus Puer.[38] Both designations, Infans and Puer, imply a young age. The former also appears in a Hereford charter of 1195, where a Master Roger Infans is called as a witness. Additional support for the identification of Roger Herefordensis with Roger Infans is provided by a set of astronomical tables drawn up by 'Roger Infans' and now held in Brussels, which are also attributed to Roger of Hereford. In addition, a now-lost document in John Dee's (1527–1608/9) library[39] is said to have contained a *computus* by Roger of Hereford, which was also attributed to Roger Infans.[40]

In the preface to his *Computus*, dated 9 September 1176, Roger describes himself as 'a young man who seems presumptuous in rehandling so many writings of the ancients'. Yet, he also claims to already have devoted many years of 'sweaty' (*desudavi*) hard work to the *regimen scholarum*, the scholarly regime. Perhaps falsely modest, he adds that he realises how difficult it will be to add anything substantial to a subject that has attracted attention, *saepe et diligenter* (often and careful), from the era's most celebrated scholars. Nevertheless, he points out that he was driven to compose his treatise 'at the request of many who were attracted by the excellence of this science'.[41] The treatise is dedicated to Gilbert—*Gilleberto Rogerus salutes H[ic?] D[icit?]* ('*Gilbert, Roger greets here*'). The person of interest here is most likely Bishop Gilbert Foliot (ca. 1110–1187), in whose household Roger spent a considerable time[42] and whom Roger considered his patron.[43] By 1176, Foliot had moved on from his earlier appointment as Bishop of Hereford (1148–1163) to become Bishop of London (1163–1187).

Roger's alternative Latin name, Infans or Puer, was anglicised by the historian John Leland (1506–1522) to 'Yonge' (also 'Young'), whereas John Wallis (1616–1703), in his translation of *A Treatise of Algebra* (1693), first refers to him as 'Roger Child(e)'. On occasion, we also find references to Roger of Hereford's youth using the Anglo–Norman designation L'Enfant or Lenfant. Roger's name ambiguity has resulted in duplicate entries in the *Dictionary of National Biography*, including 'Roger Infans (fl. 1124)' and 'Roger of Hereford (fl. 1178).' The earlier era of activity implied by the former entry, initially composed by Thomas Wright in 1842, has mistakenly taken the number 1124 from the margin of Roger's *Computus*, where it represented an intermediate result of a calculation devoted to astronomical discrepancies rather than the year of composition.[44] The date 1176 appears clearly in two other passages in the same manuscript.

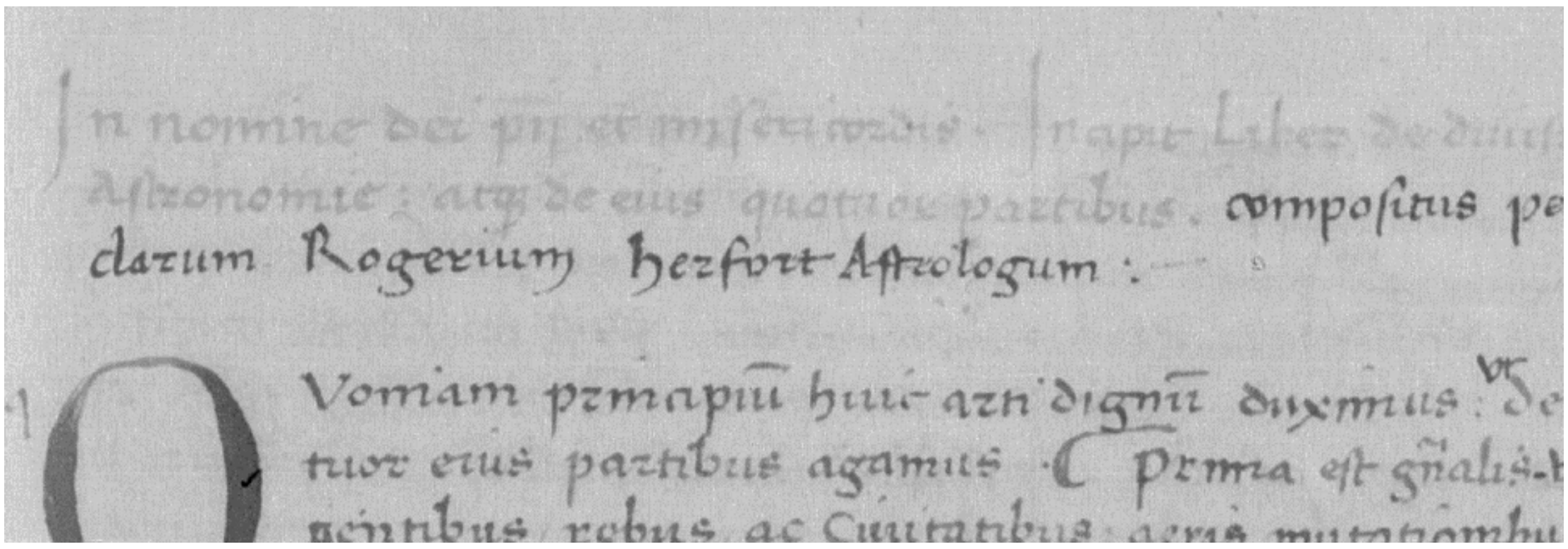

**Figure 3.** *Incipit* (manuscript opening) with an 'Islamic' dedication: *In nomine dei pii misericordis* (In the name of the merciful God) ... *Compositus per clarum Rogerium Hertford Astrologum.* (Paris, Bibliothèque nationale de France, Lat. 10271, f.179r; public domain)

The earliest record of Roger of Hereford dates from a commission, in 1176, by Bishop Foliot of a *computus* for the calculation of ecclesiastical dates. However, the *Dictionary of National Biography* suggests (without providing any evidence or provenance) that he was a laborious student at Cambridge,[45] held in high esteem by his peers. This latter assertion is supported by the fact that his friend, Alfred of Sareshel, 'the Englishman', translated the Arabic version of the pseudo-Aristotelian manuscript *De vegetabilibus* (attributed to Nicolaus Damascenus; ca. 40 B.C.E.), dedicating it to '*magistro Rogero de Herfodia*'. Since the University of Cambridge was not established until 1209, it is likely that he received his ecclesiastical education from the monks at the nearby bishopric church of Ely. Roger developed into a scholar of the 'astronomical arts'[46]—*viz.* astronomy, astrology and the ecclesiastical *computus*—and became known as an expert in natural philosophy, alchemy (mines and minerals) and mathematics.

It is not clear whether he ever travelled abroad,[47] nor whether he could read or understand Arabic,[48] yet his work was clearly influenced by Arabic science.[49] For instance, one of Roger's astronomical treatises, the *Theorica planetarum Rogeri Herefordensis*,[50] has a characteristically Islamic opening (see Figure 3): 'In the name of God the pious and merciful, here opens the book of the division of astronomy and its four parts composed by the famous astrologer Roger of Hereford.' This particular manuscript discusses the properties of planets in different countries, including Arabia, Turkey, India, Babylonia and Spain. It describes the 'Hindu' procedure to determine planetary latitudes, a technique that was most likely introduced by Arab scholars. The treatise also provides the 'more likely' Ptolemaic method for latitude determination. Although this is not the place to delve into the differences between both methods, the fact that Roger references both techniques and undertakes a critical assessment of their operation demonstrates the prevailing tension between the ancient and new approaches to learning in the late twelfth century.

Roger's date of birth is unknown, and so based on his own reference to his youth in his 1176 *Computus*, it appears that he may have been born some time in the 1150s. In this context, Harvard University's Charles Haskin's identification[51] of a 'Master Roger of Hereford' attesting a York charter of 1154–1163 appears too early. In a broader context, Josiah C. Russell (University of New Mexico) laments, 'The chief difficulty seems to come from the popularity of the name Roger at Hereford.'[52]

Nevertheless, it seems clear that Roger spent a significant fraction of his career writing and teaching as Clerk of Hereford. He was clearly more enamoured with the new Arabic science ('*Arabum studia scrutarum*': Arabic enthusiasms) than most of his contemporaries. His intellectual circle included Alfred of Sereshal and, most likely, Alexander Neckham (1157–1217), the English magnetician, poet, theologian and writer. In addition, the Pipe Rolls (financial records maintained by the English Exchequer) reveal[53] that he successfully used his strong and wide connections across the higher levels of society—including his powerful patron, Henry II of England (1133–1189; reigned from 1154)—to obtain an appointment as itinerant justice for Herefordshire, Gloucestershire, Worcestershire, Shropshire and Staffordshire in 1185–1186,[54] together with Walter Map.

### *Computus*

Roger of Hereford produced a number of groundbreaking manuscripts during his professional career, including his celebrated *Computus* of 1176. Development of the science of *computus* in Western European cathedral schools in the eleventh and twelfth centuries was driven by the Christian need to calculate the date of Easter, for which accurate predictions of the solar and lunar cycles were required. It was grounded in ideas originating in the (much earlier) Celtic church, combined with influences from Lotharingian and Jewish scholars.[55]

Roger wrote his *Computus* to achieve ecclesiastical calendar reform. He made explicit use of the new Arabic and Greek science to mitigate the prevailing calendrical problems.[56] Those problems were predominantly caused by inaccurate astronomical measurements.[57] As we will see below, Roger's improved ecclesiastical calendar was hence based not only on the traditional approaches usually associated with the widely available and popular *computus* of Dionysius Exiguus, the 'Humble'—the sixth-century Scythian monk best known for his invention of the 'Anno Domini' dating—but it also took into account observed astronomical phenomena.[58]

Roger's *Computus* comprises five books, which are in turn divided into 26 chapters.[59] In the preface to his treatise, he criticises the teaching of astronomy in the European cathedral schools. He is particularly scathing about the calculations of Gerland, Helferic and other Latin computists, lamenting their conservatism and their ignorance of Arabian astronomy. Instead, he proceeds by adopting Hebrew and Chaldean (Arabic) sources in order to compose his own, improved *Computus*. Gerland's earlier *Computus* appears to have been more incremental than truly innovative. The scholar merely proposed an alternative date for the Incarnation of Christ, but he did not attempt to correct the intrinsic weaknesses of the prevailing ecclesiastical reckoning.

The most important shortcoming of the ecclesiastical calendars in use in the eleventh and twelfth centuries was their reliance on the so-called 19-year cycle (*ciclus decennovenalis*) in the Dyonisian reckoning, which equated 19 solar years with 235 lunations (lunar months).[60] A key problem associated with its use was that the 19-year cycle does not repeat *ad infinitum*. Although the ecclesiastical calendar's main shortcomings were well-known by the eleventh century,[61] the church calendar carried the Council of Nicaea's stamp of approval. Roger is reported to have stated,[62] 'We dare not change anything relating to the lunations of the ecclesiastical *computus*', given the Nicaean bishops' edict.[63]

Roger nevertheless attempted to enact reform of the ecclesiastical calendar's reliance on the 19-year cycle by incorporating calculations based on the so-called 'golden number'. The golden number was a device originating from the Roman calendar—possibly from Celtic traditions—that was used to mark all of the New Moons ('primations') on the calendar throughout the 19-year cycle. It was, hence, practically useful to determine the Moon's age at any given time. Roger realised, however, that even with this improvement, the calendar tables could not be repeated in perpetuity. The main problem associated with the application of the golden number was that the lunar measurements used for the calendar tables were simply too inaccurate, in the sense that there was a tendency to overestimate the length of the lunations. In turn, that observational inaccuracy resulted, over time, in a gradually increasing discrepancy between the calculated and actual dates of the New Moon. Roger's proposed calendar reform was based on precisely that realisation.

Book 5 of Roger's *Computus* clarifies that his revised calendar tables, his 'natural *computus*', start from an observed astronomical event, a solar eclipse that occurred in Gerland's time, on 23 September 1086 at 15:00 hours—or, according to the prevailing Dyonisian reckoning, in 1093. That time corresponds to 09:00 Greenwich Mean Time (GMT), since according to 'divine authority' (*Divina autem auctoritas*) a given day commenced at the start of the previous evening, that is, at 18:00 GMT.[64]

Despite Roger's assertion that his *Computus* was based on an actual astronomical event, he conceded that his tables were still based on those of Gerland,[65] and that any adjustments were entirely computational. His contemporaries were not universally prepared to accept Roger's calendar reform, particularly not because adoption of Arabic science was slow and inhomogeneous across northwestern Europe. In addition, a theological debate was raging between proponents of the 'natural *computus*', which was based on reason ('a vain and empty science not visible to the eyes or audible to the ears'[66]), and those in favour of adoption of the 'vulgar *computus*', which relied on the senses, but which was also seen as lacking subtlety. Embroiled in this ongoing controversy, Roger attempted nothing less than correction of the prevailing calendrical errors by substituting traditional, 'natural' reckoning for reformed, 'vulgar' reckoning. He was clearly far ahead of his time.

**The Hereford meridian**

Roger continued to work on solving astronomical problems for Bishop Foliot.[67] Two years after the completion of his *Computus*, he embarked on a venture that would eventually link his name inextricably to Hereford. In 1178, Roger calculated the conversions required to adapt the existing Arabic astronomical and astrological tables for Toledo and Marseille to the longitude of Hereford.[68] Most Latin versions of the Toledo and Marseille Tables[69] included appendices containing lists of cities across the continent and their astronomical coordinates.[70] These might be used to transpose the tables to a range of different longitudes and latitudes.

Astronomical and astrological tables were first compiled by the Persian astronomer al-Khwārizmī. Adelard of Bath and, subsequently, one Roger of Chester[71] worked with slightly revised and improved versions of al-Khwārizmī's tables. Roger of Chester produced a Latin translation in 1145.[72] During the twelfth century, al-Khwārizmī's tables were gradually replaced by a revised set of tables and an accompanying commentary (the *Canones*; see Figure 4),[73] which had been produced specifically for Toledo by the Spanish–Arab (Andalusian) astronomer Abū Ishāq Ibrāhīm al-Zarqālī (1029–1100) in 1080. In 1141, the Toledan Tables were revised and translated into Latin by Raimond (Raymond) de Marseille,[74] who may have taken them to England shortly afterwards.

The Toledan Tables were used to calculate the equivalent tables for other locations in France, England, Italy and possibly also elsewhere across Europe.[75] They offered the tantalising prospect of accurately predicting planetary positions. This represented the overarching desire of medieval astronomers; their interest in longitudes and latitudes was predominantly astronomical (astrological) rather than geographical.[76] Contemporary tables were usually accompanied by instructions how to transpose solar, lunar, stellar and other celestial positions based on the meridian of a reference city to the longitudes of others. Such instructions[77] often involved the use of solar or lunar eclipse observations in order to allow calculations of longitude differences with respect to the reference location.

Other suggested approaches included a rudimentary version of the 'lunar distance' method routinely used for longitude determination at sea.[78] In his thirteenth-century *Theorica planetarum*[79] Gerard of Cremona writes,

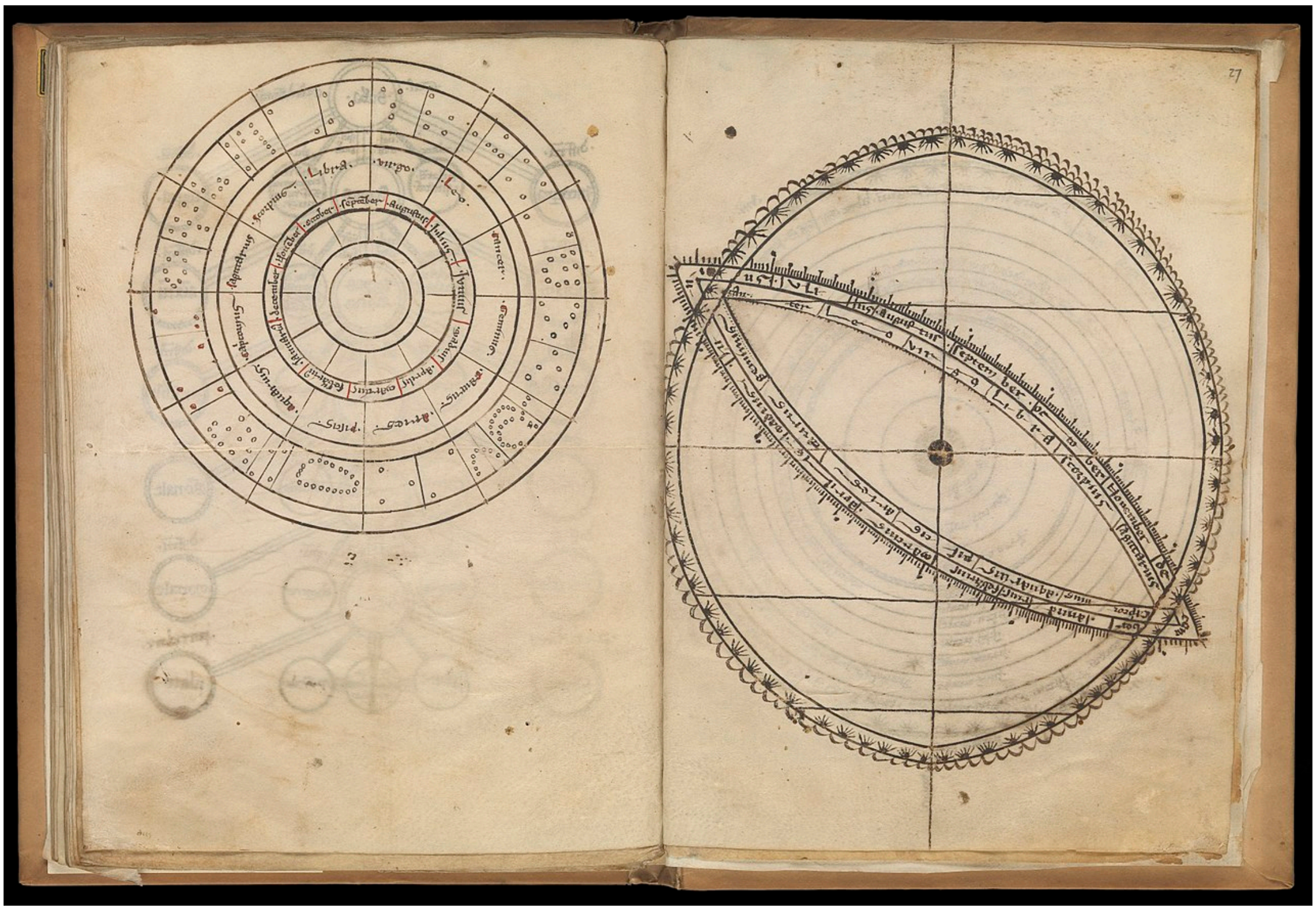

**Figure 4.** The *Canones* are a translation of a diagram from the Toledan Tables by Gerard of Cremona. (Courtesy: Wellcome Collection, L0070081; Creative Commons Attribution-Share Alike 4.0 International license)

> When the Moon is on the meridian, if you compare her position with that given in the lunar tables for some other locality, you may determine the difference in longitude between the place where you are and that for which the lunar tables were constructed by noting the differences in the position of the Moon as actually observed and as recorded in the tables. It will not be necessary for you to wait for an eclipse.[80]

It is unclear whether this approach was, at least at the time, ever successfully applied in practice. We do know, however, that Roger of Hereford used the lunar eclipse of 12 September 1178 to simultaneously determine the meridian differences between Toledo, Marseille and Hereford with respect to the longitude of Arin, the mythical centre of the Islamic world.[81] This placed Hereford firmly on the map. Roger's Hereford tables also included the longitudes and latitudes of more than 70 other locations, many of them in the Islamic world. This represented a significant improvement with respect to extant lists of geographical positions.[82]

Roger used a version of al-Khwārizmī's tables that had been translated and revised by Robert of Ketton, the English astronomer, translator and diplomat based in Spain. These complemented the tables calculated by Raimond de Marseille, with which Roger was clearly familiar. He is known to also have relied on Raimond in some of his other works, 'as well as on translations of Arabic astrological texts made by Juan de Sevilla and Hermann of Carinthia'.[83] He chose to use the Christian calendar as the basis for his meridian calculations, 'because the years of the Arabs and their months are difficult to our people who are not accustomed to them'.[84]

The practice of taking advantage of naked-eye lunar eclipse observations to determine longitude differences had already been well established by the time Roger obtained his own observations and determined the Hereford meridian in 1178. The theological scholar and astronomer Walcher, Prior of Great Malvern (d. 1135), obtained fairly reliable results from simultaneous observations in England and Italy of the lunar eclipses of 30 October 1091[85] and 18 October 1092, and of some of those occurring in 1107–1112.[86] Walcher brought the astrolabe to England as early as 1091 or 1092. In fact, his use of the astrolabe to record the exact time of the lunar eclipse of 18 October 1092 was the first recorded use of the instrument in England. He referred to two of its points by their Arabic names, presumably convinced that those designations were well-known by his intended audience.

Walcher was among the first to embrace the teachings of Petrus Alphonsi (Anfusi; d. after 1116), a converted Spanish (Aragonese) Jew and considered an important scholar in the dissemination of Arabic astronomy to England. For instance, Alphonsi's *Dialogi cum ludeo* (also known as the *Dialogus* or the *Dialogue against the Jews*; see Figure 5) offers a clear exposition of the principles of local time differences.[87] Walcher gained a reputation as a mathematician and astronomer, with a firm grounding in Arabic science and mathematics.

Between 1107 and 1112 he published a set of lunar tables containing the times of mean conjunctions, calculated in a simple scheme,[88] with explanations. These comprised a cycle of 76 years leading up to 1112, all calibrated based on the eclipse of 1092. The calculations published in his first treatise were based on application of awkward Roman fractions, but by 1120 he had moved on to using the degrees, minutes and seconds, and the more exact observations he had learnt from Petrus Alphonsi. Mindful of Alphonsi's teachings and inspired by al-Khwārizmī's insights, he thus proceeded to compose a treatise for predicting eclipses, *De Dracone* (1120–1121).[89]

## Concluding thoughts

As we have seen, Roger of Hereford propelled Hereford into the scholarly and ecclesiastical limelight of the late twelfth century by embracing the 'new science' from Arabic sources well before most of his contemporaries. As such, he put the cathedral city quite literally on the map. Hereford thus joined the ranks of other illustrious seats of learning across the European continent for which individual meridian tables were computed in rapid succession—Toledo and Marseille, London,[90] Toulouse,[91] Cremona[92] and Novara,[93] as well as Paris, Palermo, Pisa, Constantinople and Genoa.[94]

Hereford's inclusion among the leading centres of learning in the twelfth century did not occur by accident. Its strategic location at the physical boundary between England and Wales and at the virtual crossroads between traditional theological doctrines and the new Arabic science, supported by the enlightened leadership of Bishops Robert of Lorraine and Foliot, attracted eminent scholars from far and wide. Roger of Hereford's residence as a teacher at the cathedral school acted as a conduit between the old and new scholarship, between Adelard of Bath, Petrus Alphonsi, Robert Grosseteste and the ancient authors.

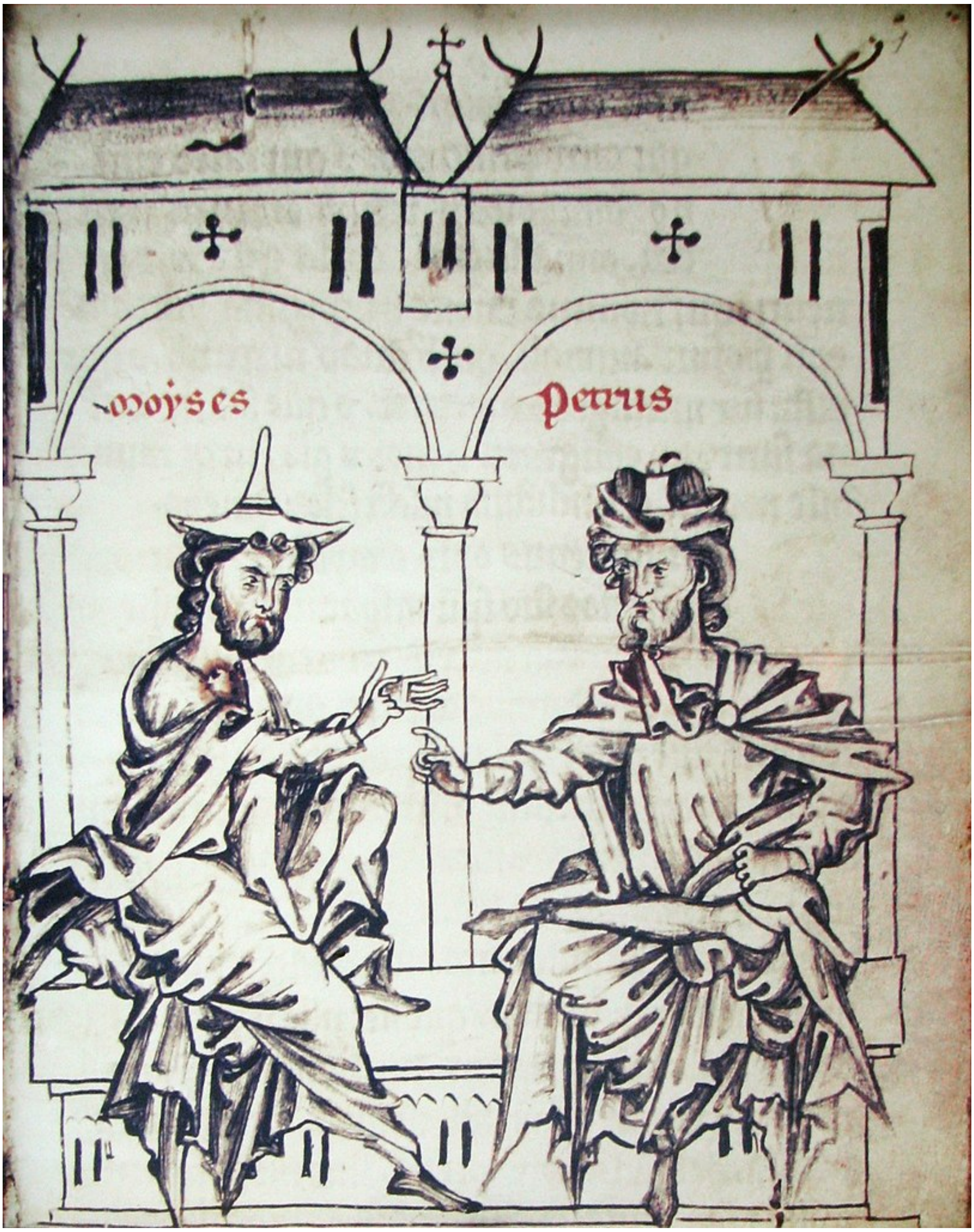

**Figure 5.** Excerpt from the *Dialogi contra Iudaeos* by Petrus Alphonsi, illustrating a dialogue between the Jew *Moyses* and the Christian *Petrus*. (Grootseminarie Ten Duinen, Bruges, Manuscript 26/91, via Wikimedia Commons; public domain)

Roger's innovations significantly advanced the study of mathematical sciences in England. He also invented a new way of calculating horoscopes mathematically.[95] In fact, his strong network provided him with the opportunity to cast Eleanor of Aquitaine's (ca. 1122–1204; Queen of France and wife of Henry II) natal horoscope.[96] In the preface to his *Computus*, Roger implies that he had taught several years, although he does not provide any details as to where he had spent that time. The widespread availability of his *Liber de Astronomice iudicandi* (*Judicial Astrology*) suggests that this manuscript may have served as a teaching document.[97] In addition, if he had prepared his astronomical tables for educational purposes, their use of the Hereford meridian as reference may indeed be significant.[98]

Nevertheless, of great significance for the progress of philosophical developments in England and the central role played by Hereford in the dissemination of Arabic innovations was the association, both personal and professional, of Roger with Alfred of Seneshal. The latter's work marks an important milestone in the intellectual development of medieval England. It clearly went beyond the mathematical and astronomical concerns on which Roger focused, but tentatively expanded into the philosophical currents of Oxford at the beginning of the thirteenth century.[99]

In conclusion, the fortuitous presence of a number of enlightened individuals, combined with a strategic location and politically favourable conditions, allowed the cathedral city of Hereford to play a leading intellectual role on the world stage in the late twelfth century. Such developments would likely have happened at some time during the late medieval era, irrespective of the scholars involved. However, the adoption of the new Arabic science by Roger of Hereford acted as an early stimulant, thus cementing Hereford's fame as a progressive force in the annals of history.

**Acknowledgments**
I thank Dr. Doru Costache (Sydney College of Divinity and Institute for the Study of Christianity in an Age of Science and Technology) for his critical reading and constructive criticism of a close-to-final version of this article. I also thank Roz Lowe and Terry Morgan of the Woolhope Club, Herefordshire, for sending me a scanned copy of the French (2000) reference.

NOTES AND REFERENCES

[1] Following the usual convention, 'Arab', 'Arabic' and 'Arabian' loosely refer to, respectively, the people, language and geography of the region.

[2] Dorothee Metlitzki, *The matter of Araby in medieval England* (Yale University Press, 1977) 30

[3] Chris Mitchell, *Roger of Hereford's* Judicial Astrology*: England's First Astrology Book?* (PhD thesis University of Leicester, 2019)

[4] Around the year 833, Muḥammad ibn Mūsā al-Khwārizmī (ca. 780–ca. 850), the Persian mathematician, astronomer, and geographer in the House of Wisdom in Baghdad, compiled his *Book of the Depiction of the Earth*, which closely resembled Ptolemy's *Geographia*; Yossef Rapoport and Emilie Savage-Smith, 'The Book of Curiosities and a Unique Map of the World', in Richard J. A. Talbert and Richard W. Unger (eds) *Cartography in Antiquity and the Middle Ages: Fresh Perspectives, New Methods* (Koninklijke Brill NV, 2008) 121–138

[5] C. Raymond Beazley, *The Dawn of Modern Geography* vol.I: *A History of Exploration and Geographical Science from the Close of the Ninth to the Middle of the Thirteenth Century (c. AD 900–1260)* (John Murray, 1897); Keith D. Lilley, 'Geography's medieval history: A neglected enterprise?', *Dialogues in Human Geography* vol.1(2) 147–162

[6] Valentin Rose, 'Ptolemaeus und die Schule von Toledo', *Hermes* vol.viii (1874) 327–349; Charles H. Haskins, 'The Reception of Arabic Science in England', *The English Historical Review [EHR]* vol.XXX (1915) 56–69

[7] Nicholas Whyte, *Roger of Hereford's* Liber de Astronomice iudicandi*: A Twelfth-Century Astrologer's Manual* (MPhil thesis University of Cambridge, 1991)

[8] John K. Wright, 'Notes on the knowledge of latitudes and longitudes in the middle ages', *Isis* vol.5(3) (1923) 75–98

[9] Roger French, 'Foretelling the Future: Arabic Astrology and English Medicine in the Late Twelfth Century', *Isis* vol.87(3) (1996) 453–480

[10] Matthew F. Dowd, *Astronomy and compotus at Oxford University in the early thirteenth century: The works of Robert Grosseteste* (PhD thesis University of Notre Dame, 2003)

[11] Haskins, *EHR*, 56

[12] George L. Hamilton, 'Langlois, Charles-V., La Connaissance de la Nature et du Monde au Moyen Age', *Romanic Review* vol.III (1912) 314–434

[13] Philippe de Thaon and Eduard Mall (ed), *Li cumpoz Philippe de Thaün: Der Computus des Philipp von Thaun* (K. J. Trübner, 1873)

[14] Melissa Petruzzello and The Editors of Encyclopaedia Britannica, 'St. Bede the Venerable: Anglo–Saxon historian', *Encyclopaedia Britannica*, https://www.britannica.com/biography/Saint-Bede-the-Venerable [accessed 30 June 2022]

[15] Steven J. Livesey and Richard H. Rouse, 'Nimrod the Astronomer', *Traditio* vol.37 (1981), 203–266

[16] British Library, Sloane MS 263, *Helperic of Auxerre*

[17] Haskins, *EHR*, 56, note 3

[18] I will use the spelling *computus* ('calculation') when referring to the science of *computus* and to generic manuscripts in this same context. I note, however, that late medieval ecclesiastical notation gradually changed to *compotus* ('account').

[19] W. E. van Wijk (ed), 'Le comput emendé de Reinherus de Paderborn (1171)', *Verhandelingen der Koninklijke Nederlandse Akademie van Wetenschappen (afd. Letterkunde)*, vol.57 (1951)

[20] Digby MS 40, fol.22v; own translation

[21] Moritz Steinschneider, 'Abhandlung zur Geschichte der mathematischen Wissenschaften im XII. Jahrhundert', *Zeitschrift für Mathematik und Physik* vol.25 (1880) 57–128; Joseph Jacobs, *The Jews of Angevin England* (G. P. Putnam and Sons, 1893); Haskins, *EHR*, 68

[22] Alan B. Cobban, *The Medieval English Universities: Oxford and Cambridge to c. 1500* (University of California Press, 1988) 27–28

[23] Reginald L. Poole, *The Exchequer in the Twelfth Century, the Ford Lectures Delivered in the University of Oxford in Michaelmas Term 1911* (Trieste Publishing Pty Ltd, 2017) 47ff; Charles H. Haskins, 'The Abacus and the King's Curia', *EHR* vol.XXVII(CV) (1912) 101–106; Charles Burnett, 'Mathematics and Astronomy in Hereford and Its Region in the Twelfth Century', in David Whitehead (ed) 'Medieval Art, Architecture and Archaeology at Hereford', *British Archaeological Association Conference Transactions* vol.15 (1995) 50–59

[24] The earliest evidence of a Jewish community in Hereford dates from 1178–1179; Jennifer Moreton, 'Before Grosseteste: Roger of Hereford and Calendar Reform in Eleventh- and Twelfth-Century England', *Isis* vol.86(4) (1995) 585

[25] Lotharingian science was disseminated in England by Abbo of Fleury (ca. 945–1004), who taught astronomy and *computus* to the monks of Ramsey from 986 to 988; Moreton, *Isis*, 562–586

[26] J. W. Thomson, 'The Introduction of Arabic Science into Lorraine in the Tenth Century', *Isis* vol.12 (1929) 184–193; Alfred Cordoliani, 'L'activité computistique de Robert, évêque de Hereford', in Pierre Gallais and Yves-Jean Rious (eds), *Mélanges offerts à René Crozet*, vol.1 (Société des Études Medievales, 1966), 333–334

[27] Josiah C. Russell, 'Hereford and Arabic Science in England about 1175–1200', *Isis* vol.18 (1932) 14–25; Adrian Morey and Christopher N. L. Brooke, *Gilbert Foliot and His Letters* (Cambridge University Press, 1965) 197

[28] Oxford, Bodleian Library, *Prefatio magistri Rogeri infantis in compotum*, Digby MS 40, s.xii–xiii, fols. 21–50v; London, British Library, *Roger of Hereford, Astrological Tables*, Arundel MS 377, ff. 77–85; Cambridge, University Library, *Ptolemaeus Arabus et Latinus. Manuscripts*, MS Kk 1.1

[29] Metlitzki, *The matter of Araby in medieval England*, 59; Thomas Wright, *Biographia Britannica Literaria (1842)* (Read Books, 2012) 90–91

[30] Roger French, 'Roger of Hereford: Astrology and the School of Hereford', in David Whitehead and John Eisel, (eds), *A Herefordshire Miscellany: Commemorating 150 Years of the Woolhope Club* (Lapridge Publications, 2000) 247–255

[31] Metlitzki, *The matter of Araby in medieval England*, 19; for the full poem, see Public Record Office (Great Britain), *Rerum Britannicarum Medii Aevi Scriptores* (*Chronicles and Memorials of Great Britain and Ireland During the Middle Ages*) vol.21(1) (Longman, Brown, Green, Longmans and Roberts, 1861) 382–384

[32] Translation: Jack P. Cunningham and Steven Puttick (eds), *Robert Grosseteste and Theories of Education: The Ordered Human* (Routledge International Studies in the Philosophy of Education, 2019) (no page numbers)

[33] French, *A Herefordshire Miscellany* 247–255

[34] Dowd, *Astronomy and compotus at Oxford University*, 37; French, *A Herefordshire Miscellany* 247–255

[35] Jack P. Cunningham and Mark Hocknull (eds), 'Robert Grosseteste and the pursuit of Religious and Scientific Learning in the Middle Ages', *Studies in the History of Philosophy of Mind* vol.18 (Springer,

2016); Greti Dinkova-Bruun, Giles E. M. Gasper, Michael Huxtable, Tom C. B. McLeish, Cecilia Panti and Hannah Smithson, 'The dimensions of colour: Robert Grosseteste's *De colore*', *Durham medieval and renaissance texts* vol.4 (Pontifical Institute of Mediaeval Studies, 2013); Tom C. B. McLeish, Richard G. Bower, Brian K. Tanner, Hannah E. Smithson, Cecilia Panti, Neil Lewis, Giles E. M. Gasper, 'History: A medieval multiverse', *Nature* 507 (2014) 161–163

[36] This reference to Hereford possibly reflects that Roger owned land in Sufton, in the parish of Mordiford; French, *A Herefordshire Miscellany*, 247–255

[37] Oxford, Bodleian Library, Digby MS 40

[38] Russell, *Isis*, 15

[39] Julian Roberts and Andrew G. Watson (eds), *John Dee's Library Catalogue* (The Bibliographical Society, 1990)

[40] Whyte, *Roger of Hereford's* Liber de Astronomice iudicandi

[41] Metlitzki, *The matter of Araby in medieval England*, 39; Th. Wright, *Biographia Britannica Literaria*, 90

[42] Charles H. Haskins, *Studies in the History of Mediaeval Science* (Harvard University Press, 1924) 124–125

[43] Whyte, *Roger of Hereford's* Liber de Astronomice iudicandi

[44] Jacqueline A. Stedall, 'Of Our Own Nation: John Wallis's Account of Mathematical Learning in Medieval England', *Historia Mathematica* vol.28 (2001), 73–122

[45] William F. Sedgwick, 'Roger of Hereford', *Dictionary of National Biography, 1885–1900* vol.49 (2020) https://en.wikisource.org/wiki/Dictionary_of_National_Biography,_1885-1900/Roger_of_Hereford [accessed 27 June 2022]

[46] French, *A Herefordshire Miscellany*, 247–255

[47] It has been suggested that he gained a familiarity with Arabic treatises and proficiency in their translation in Spain, where he may have contributed actively and systemically to translation work, in a community of Latin, Mozarab and Jewish scholars, both at Toledo and at other seats of learning in the Ebro valley and the Pyrenees; Metlitzki, *The matter of Araby in medieval England*, 30

[48] Dowd, *Astronomy and compotus at Oxford University*

[49] Russell, *Isis*, 14–25; French, *Isis*, 453–480

[50] Oxford, Bodleian Library, *Theorica Planetarum Rogeri Herefordensis*, Digby MS 168

[51] Russell, *Isis*, 15

[52] *ibid.*

[53] Pipe Roll 31, *Henri II* (1185) 146; Haskins, *Studies in the History of Mediaeval Science*, 126

[54] Whyte, *Roger of Hereford's* Liber de Astronomice iudicandi

[55] Moreton, *Isis*

[56] Roger French, 'Teaching Aristotle in the Medieval English Universities: *De plantis* and the *Physical Glossa ordinaria*', *Physis* vol.34 (1997) 225–296

[57] Moreton, *Isis*; Dowd, *Astronomy and compotus at Oxford University*, ch.4; Olaf Pedersen, 'The *Corpus astronomicum* and the Traditions of Medieval Latin Astronomy', *Studia Copernicana* vol.3 (1973) 57–96; Jennifer Moreton, 'Sacrobosco and the Calendar', *Viator* vol.25 (1994) 229–244

[58] Moreton, *Isis*

[59] Digby MS 40, ff. 21–50v

[60] Charles W. Jones (ed), *Bedae Opera de temporibus* (Medieval Academy of America, 1943) 3–104; Moreton, *Isis*

[61] W. E. van Wijk, *Verhandelingen*, liv.II, Ch.1ff

[62] Note that the Nicaean canons might have been mis- or spuriously quoted in late medieval England as much as other Greek sources, e.g., works attributed to Saint John Chrysostom; Daniel Anlezark, 'The Reception of John Chrysostom in Early Medieval England', in Doru Costache and Mario Baghos (eds), *John Chrysostom: Past, Present, Future* (AIOCS Press, 2017) 71–85

[63] Digby MS 40, fol.48r

[64] Moreton, *Isis*; Jones, *Bedae Opera de temporibus*, 189; Bradley E. Schaefer, 'Astronomy and the Limits of Vision', *Vistas in Astronomy* vol.36 (1993) 311–361

[65] Digby MS 40, fol.49v

[66] Moreton, *Isis*, 573

[67] Based on his broad expertise in natural philosophy, astronomy, astrology, mines and alchemy (mines and minerals), he composed six major works: (*i*) *Theorica Planetarum Rogeri Herefordensis* (Oxford, Bodleian Library, Digby MS 168), (*ii*) *Introductorium in artem judiciariam astrorum*, (*iii*) *Liber de quatuor partibus astronomiæ judiciorum editus a magistro Rogero de Herefordia* (Digby MS 149), (*iv*)

*De ortu et occasu signorum*, (*v*) *Collectaneum annorum omnium planetarum* and (*vi*) *De rebus metallicis* (Peterhouse College, Cambridge; now lost).
[68] Madrid MS 10016, *Astronomical tables for the meridian of Hereford in 1178, based upon tables for Toledo and Marseilles*, ff. 4, 73–83v, 85; British Library, Arundel MS 377, ff. (77–85), 86v–87: *Anni collecti omnium planetarum compositi a magistro Rogero super annos domini ad mediam noctem Herefordie anno ab incarnatione domini .m°.c°.lxx°.viii°. post eclipsim que contigit Hereford eodem anno (13 September)*. It has been suggested that Roger revised his tables in 1184; Whyte, *Roger of Hereford's* Liber de Astronomice iudicandi
[69] Gerald J. Toomer, 'A Survey of the Toledan Tables', *Osiris* vol.15 (1962) 5–174
[70] J. K. Wright, *Isis*
[71] It is not possible to identify this person beyond any reasonable doubt given the popularity of this name, combined with his relatively modest scholarly contributions.
[72] Madrid, Biblioteca Nacional, *Incipit liber Ezeig id est chanonum Alghoarizmi per Adelardum Bathoniensem ex arabico sumptus et per Rodbertum Cestrensem ordine digestus*, MS 10016, f.8
[73] Pierre Duhem, *Le système du monde; histoire des doctrines cosmologiques de Platon à Copernic* vol.II. (Hermann, 1914) 250–251
[74] Burnett, *Medieval Art, Architecture and Archaeology at Hereford*, 55
[75] Moritz Steinschneider, 'Etudes sur Zarkali', *Bullettino di bibliografia e di storia delle scienze matematiche e fisiche* (B. Boncompagni) vol.XIV, 1881; vol.XVI, 1883; vol.XVII, 1884; vol.XVIII, 1885; vol.XX, 1887
[76] J. K. Wright, *Isis*
[77] Paris, Bibliothèque nationale, MS, Fonds Latin, n° 14704, fol.116, col.c
[78] Richard de Grijs, 'A (not so) brief history of lunar distances: Lunar longitude determination at sea before the chronometer', *Journal of Astronomical History and Heritage* vol.23 (2020) 495–522
[79] Olaf Pedersen, 'The Origins of the Theorica Planetarum', *Journal for the History of Astronomy*, Vol.12 (1981) 113–123
[80] Translation: J. K. Wright, *Isis*, 83–84
[81] *ibid*., 85
[82] *ibid*., 87–88
[83] Charles Burnett, 'Roger of Hereford', in Thomas Hockey et al. (eds), *Biographical Encyclopedia of Astronomers* (Springer, 2007) 982
[84] Th. Wright, *Biographia Britannica Literaria*, vol.II, 219; W. E. van Wijk, *Verhandelingen*
[85] While travelling in eastern Italy, Walcher had not been able to measure the exact timings, shortly before dawn, of the lunar eclipse of 30 October 1091, but he noted upon his return to England that a fellow monk had observed the same eclipse at a considerably different time of day, just before local midnight; J. K. Wright, *Isis*, 81; Haskins, *EHR*, 57
[86] John K. Wright, 'Notes on the knowledge of latitude and longitude in the Middle Ages', *Isis* vol.5 (1922) 75–98; Armando Cortesão, *History of Portuguese Cartography* (1. Junta de Investigações do Ultramar, 1969) 182–183
[87] Haskins, *EHR*, 56, 57, 61; J. K. Wright, *Isis*, 82–83
[88] Walcher of Malvern, *De lunationibus* (*Lunar tables 1036–1111, calculated from an eclipse observed on 19 Oct. 1092*; fol. 90), Oxford, Bodleian Library, MS Auct. F. 1. 9, ff.86–99
[89] Matthew S. Lampitt, 'Networking the March: A History of Hereford and its Region from the Eleventh through Thirteenth Centuries', *Mortimer History Society Journal* vol.1 (2017) 55–72
[90] Paris, Bibliothèque nationale, MS Fonds Latin, n° 7272, fol.60 col. *a*–fol.67 col. *d*; Duhem, *Le système du monde* vol.III, 231
[91] Paris, Bibliothèque nationale, MS Fonds Latin, n° 16658, fol.71vo. This manuscript contains the Toledo Tables; on folio 71vo it includes tables for 'Tolosa'.
[92] Paris, Bibliothèque nationale, MS Fonds Latin, n° 16211, fol.41ro. As for the Toulouse tables, the Toledo Tables include tables for Cremona; J. K. Wright, *Isis*, 84–85 (note 8).
[93] Paris, Bibliothèque nationale, MS Fonds Latin, n° 7406, fol.98ro. As for the Toulouse tables, the Toledo Tables include tables for Novaro; J. K. Wright, *Isis*, 85 (note 9).
[94] Paris, Bibliothèque nationale, Preamble to the *Toledo Tables* (1232), MS Fonds Latin, n° 7272, fol.67 col. *c*; Duhem, *Le système du monde* vol.III, 231–238
[95] Whyte, *Roger of Hereford's* Liber de Astronomice iudicandi; Mitchell, *Roger of Hereford's* Judicial Astrology
[96] Whyte, *Roger of Hereford's* Liber de Astronomice iudicandi
[97] Mitchell, *Roger of Hereford's* Judicial Astrology
[98] Russell, *Isis*, 16

[99] Haskins, *EHR*, 69; Metlitzki, *The matter of Araby in medieval England*, 40

*RICHARD DE GRIJS is a professor of astrophysics at Macquarie University in Sydney, Australia. Since the publication of his monograph,* Time and Time Again: Determination of Longitude in the 17th Century *(Institute of Physics Publishing, 2017), he has gradually been building up a portfolio of history of maritime navigation essays and publications, with particular emphasis on the perennial 'longitude problem'. Richard is also a volunteer guide on Captain Cook's H.M. Bark* Endeavour *replica and the Dutch East India Company's* Duyfken *replica at the Australian National Maritime Museum, where he consults on matters related to the history of maritime navigation.*